\documentclass{iucr}
\usepackage{amsmath, amssymb}
     \paperprodcode{a000000}      
     \paperref{xx9999}            
     \papertype{FA}               

     \paperlang{english}          
     \journalcode{A}              
     \journalyr{2003}
     \journaliss{1}
     \journalvol{59}
     \journalfirstpage{000}
     \journallastpage{000}
     \journalreceived{0 XXXXXXX 0000}
     \journalaccepted{0 XXXXXXX 0000}
     \journalonline{0 XXXXXXX 0000}

\begin{document}                  



\title{Finite Size Effect of Nanoparticles to the Atomic Pair Distribution Functions}


\cauthor[ ]{Katsuaki}{Kodama}{kodama.katsuaki@jaea.go.jp,}{ }
\author[ ]{Satoshi}{Iikubo}
\author[ ]{Shin-ichi}{Shamoto}

\aff[ ]{Quantum Beam Science Directorate, Japan Atomic Energy Agency, Tokai, Ibaraki 319-1195 \country{Japan}}









\maketitle                        

\begin{synopsis}
The finite size effects of nanoparticles to the atomic pair distribution function (PDF) are calculated, which are essential 
to the structural analysis on nanoparticles.
\end{synopsis}

\begin{abstract}
The finite size effects of the nanoparticles to the atomic pair distribution functions (PDF) are 
discussed by calculating the radial distribution functions (RDF) on nanoparticles with various shapes, such as sheet, belt, 
rod, tube and sphere, assuming continua.  Their characteristics are shown depending 
on the shapes and the sizes of the 
nanoparticles.  Alternately, these PDFs can be used to measure the shapes and the sizes of ordered lattice part 
inside of any materials such as nanoparticles and bulks.
\end{abstract}



\section{Introduction}

 Many types of nanoparticles have been synthesized, and studied on 
their physical and chemical properties.  
The nanoparticles may exhibit the exotic physical properties, 
which can be different from those of the bulk materials due to their finite sizes.  
The studies on the atomic-scale structures of the nanoparticles give us information to understand their properties.
However, it is hard to determine their structures by ordinary techniques 
used for the bulk materials, for example, traditional Rietveld analysis.
In the case of the nanoparticles, the well-defined Bragg peaks are not observed 
and the diffraction data mainly consist of the diffuse scattering, because the periodicity of their unit cells is 
limited to nanoscale.   \par
The technique of the atomic pair distribution function (PDF) can be applicable to the determination of the local structure 
of the nanoparticles.
For the system composed of one kind of atom, the number of pair-atoms in 
a shell of thickness $dr$ at distance $r$ from another one are obtained as $R(r)dr$, where $R(r)$ is 
the radial distribution function (RDF).  $R(r)$ is related to the reduced 
pair distribution function $G(r)$ via the pair distribution function $g(r)$ 
as follows (Egami \& Billinge, 2003).
\begin{align}
g(r) &= R(r)/4\pi r^{2} \rho _{0}, \label{g(r)} \\
G(r)&= 4\pi r \rho _{0} [g(r)-1], \label{G(r)}
\end{align} 
where $\rho_{0}$ is a number density of atoms in the sample.
Experimentally, $G(r)$ can be obtained from 
the total scattering structure function $S(Q)$ via Fourier transformation as follows.  
\begin{equation}
G(r)=\frac{2}{\pi} \int Q[S(Q)-1]sin(Qr)dQ,
\end{equation} 
where $Q$ is the magnitude of the wave vector.  
In the cases of amorphous materials and the nanoparticles, even if $S(Q)$ does not have any well-difined peak 
as mentioned above, 
$G(r)$ can have sharp peaks, at least, in the small $r$-region.  
Then the PDF analysis is applicable to the structural analysis of the materials whose structures have 
short range correlation.  
Recently, the technique of PDF has been applied for the determination of 
the structure of nanoparticles (Gilbert \textit{et al.}, 2004, Petkov \textit{et al.}, 2004, 
Gateshki \textit{et al.}, 2004, McKenzie \textit{et al.}, 1992).\par
So far, the effect of the shape and finite size of a nanoparticle, however, has not been considered 
in the PDF analysis.  The 
atomic correlation remains only in the size, and as a result, $G(r)$ does not have any peak in the larger $r$-region 
than the particle-size if there is no correlation between near-by particles.  
Actually, $G(r)$ of C$_{60}$ has sharp peaks at $r$ 
smaller than the diameter of C$_{60}$ molecule while  
at $r$ larger than the diameter, $G(r)$ has only small broad peaks corresponding with the correlation 
between the molecules (Egami \& Billinge, 2003).  Furthermore, the intensity of 
the PDF must be reduced from that of the bulk sample with infinite size, due to the finite size 
even at $r$ smaller than the particle size.  So the distribution 
functions modified by their sizes and shapes need to be used for the detailed structural analyses of nanoparticles. \par 
Such analysis gives another structural information of the nanoparticle.  
The nanoparticle of the zinc sulfide is analyzed by PDF with the spherical shape effect  
(Gilbert \textit{et al.}, 2004). After they determined the averaged size and the shape of the nanoparticle 
by means of small angle X-ray scattering and ultraviolet-visible absorption spectroscopy, 
they discussed the disorder and the strain in the nanoparticle from the difference  
between the averaged size and the local correlation size by PDF.  The corrections of the averaged size 
are crucial to discuss the local lattice disorder in the nanoparticle.  \par
In this paper, we calculate the RDFs of the various nanoparticles by assuming 
that they are continua, for the corrections of the averaged size and the shape effect, and the correction factor to the 
RDFs and reduced PDFs for the various nanoparticles are obtained.  By using the correction factor, 
the corrected formulation of the PDF analysis on nanoparticles are presented.  Based on our calculations, 
the method to estimate the sizes of the parts with ordered and disordered lattice in the particle, is also proposed.

\section{Calculations}

In the calculation of the radial distribution function $R(r)$ of a nanoparticle, we consider 
the atomic pair distribution only in the particle, 
and assume that the atomic density is zero (vacuum) outside of the particle.
The total atomic 
density in the nanoparticle is $\rho_{0}'$.  The RDF of 
the three dimensional continuum with infinite size is given as $R_{\infty}(r)=4\pi r^{2}\rho _{0}'$, 
as mentioned in the next section.  
So the RDF of the nanoparticle, $R_{nano}(r)$, 
is modified by the correction factor $f(r)$ which is defined as, 
\begin{equation}
f(r)=R_{nano}(r)/R_{\infty}(r). \label{f(r)}
\end{equation}
When $r \rightarrow 0$, $f(r)$ must be unity, and when $r$ is larger than the size of the nanoparticle 
or $r \rightarrow \infty$, $f(r)$ 
becomes zero.  This factor can be regarded as a kind of particle form factor 
instead of a well-known atomic form factor.  \par
In the real analysis, the pair distribution between the particles must be considered.  The formulation including 
the pair distribution beween particles will be discussed in \S4.

\section{Calculated Results}
\subsection{Simple Case}

In this subsection, we take up simple cases and discuss the effect of the dimensionality of 
particles to the radial distribution function.  
First, we consider the wire with an infinitesimal thickness shown in Fig. 1(a). 
 \begin{figure}
\includegraphics{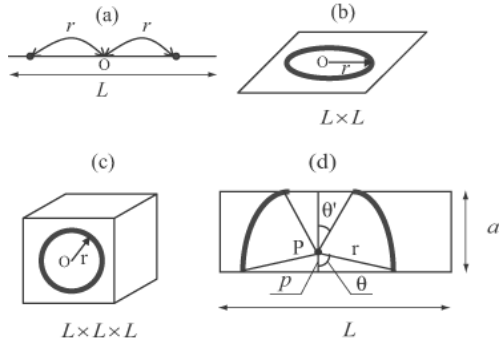} 
\caption{Schematic diagram of (a) 1D, (b) 2D, (c) 3D particles, and (d) a nanobelt.  }
\end{figure}
The density in this wire is defined as $\rho_{0, 1D}'$ whose dimension is an inverse of the length.
Here we assume that $r$ is much smaller than the length of the wire $L$.  
The number of atoms in an length $dr$ 
at the distance $r$ from the atom at origin (O), is 2$\rho_{0, 1D}' dr$, as shown in the figure.  
Then the radial distribution function is obtained as,
\begin{equation}
R_{1D}(r)=2 \rho_{0, 1D} ' .
\end{equation} 
In the case of the sheet with an infinitesimal thickness shown in Fig. 1(b), 
if $r<<L$, atoms paired with the one at origin O, 
are in the ring with a radius $r$ and a thickness $dr$ which is shown by the bold line in Fig. 1(b). 
So $R_{2D}(r)$ is given by, 
\begin{equation}
R_{2D}(r)=2 \pi r \rho_{0, 2D}  '.
\end{equation} 
In the case of the three dimensional block shown in Fig. 1(c), paired atoms are in the 
spherical shell with a radius $r$ and the thickness $dr$, and $R_{3D}(r)$ is given by, 
\begin{equation}
R_{3D}(r)=4 \pi r^{2} \rho_{0, 3D}  '.
\end{equation}
In the cases of one, two and three dimensional nanoparticles, RDF functions have the $r$-dependences of $r$-constant, 
$r$-linear and $r$-square, respectively. \par
Next, we calculate the RDF of the nanobelt with the width of $a$ and 
an infinitesimal thickness which is shown in Fig. 1(d), as an example of the typical nanoparticle.  Here, it is assumed 
that $r \ll L$.
In this case, we consider the number of atoms in an annulus of thickness $dr$ at a distance $r$ from another one 
at a position P which is distant from the edge of the belt by $p$, as shown in the figure.  
It is given by $R(r, p)dr$.  So $R(r, p)$ can be regarded as the "partial radial distribution function" (PRDF) at $p$.  
$R(r)$ is given by $\int_0^a R(r, p) dp/a$.  If 
$r \le a/2$, $R(r, p)$ is proportional to the length of the circumference for $r \le p \le a-r$, 
and it is proportional to the length of the arcs with interior angles of $2\pi-2\theta$ for $p \le r$, 
and $2\pi-2\theta'$ for $a-r \le p \le a$, where $\theta=\cos^{-1}\frac{p}{r}$ and $\theta'=\cos^{-1}\frac{a-p}{r}$.  So 
\begin{align}
R_{2Dbelt}(r)&=\left[ \int_0^r (2\pi-2\theta)r dp+\int_r^{a-r} 2\pi r dp \right. \nonumber \\
&+\left. \int_{a-r}^a (2\pi-2\theta')rdp \right] \rho_{0, 2D}'/a \nonumber.
\end{align} 
In the case that $a/2 \le r \le a$, $R(r, p)$ is proportional to the length of the arcs with interior angles 
of $2\pi-2\theta$ for $p \le a-r$ and $2\pi-2\theta'$ for $r \le p \le a$.  
At the other $p$, $R(r, p)$ is proportional to 
the length of the bold line shown in Fig. 1(d).  Then 
\begin{align}
R_{2Dbelt}(r)& =\left[ \int_0^{a-r} (2\pi-2\theta)r dp+\int_{a-r}^{r} (2\pi-2\theta-2\theta') r dp \right. \nonumber \\
& \left. +\int_r^a (2\pi-2\theta')r dp \right] \rho_{0, 2D}'/a  \nonumber.
\end{align} 
Since for $r \ge a$, $R(r, p)$ is obtained by the bold line for all $p$,
\begin{equation}
R_{2Dbelt}(r)=\left[ \int_{0}^{a} (2\pi-2\theta-2\theta' ) r dp \right] \rho_{0, 2D}'/a  \nonumber.
\end{equation} 
By calculating the above integrations, we obtain the RDF of the nanobelt as follows.
\begin{align}
R_{2Dbelt}(r)&=(2\pi a r-4r^{2}) \rho_{0, 2D}'/a  \mspace{36mu} (r\le a) \nonumber , \\
R_{2Dbelt}(r)&=\left[ 2\pi a r-4r^{2}-4ar\cos^{-1}\frac{a}{r} \right.  \nonumber \\
  & \left. +4r\sqrt{r^{2}-a^{2}} \right] \rho_{0, 2D}'/a \mspace{36mu} (r\ge a).
\end{align} 
When $r$ is much smaller than $a$, $R_{2Dbelt} \sim 2\pi r \rho_{0, 2D}'$, which is equal to $R_{2D}(r)$.  For $r \gg a$, 
$R_{2Dbelt} \sim 2a \rho_{0, 2D}'$, equal to $R_{1D}(r)$.
The calculated $R(r)$ of the nanobelt is shown in Fig. 2.
\begin{figure}
\includegraphics{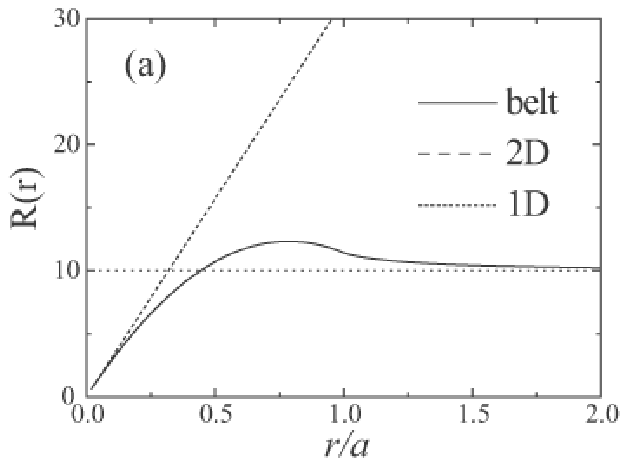} 
\caption{The radial distribution functions 
of the wire (1D), the sheet (2D) and the nanobelt. }
\end{figure}
Here, we use parameters of $\rho_{0, 2D}'=1$.  The RDFs of the wire and the sheet 
are also plotted in the figure, assuming same atomic densities.  For $r \rightarrow 0$, 
the RDF of the nanobelt approach the RDF of the 2D sheet, and for $r \gg a$, 
they merge into that of 
the wire.  We know the effect of the dimensionality to the RDF by such simple calculations. \par

\subsection{Nanosheet}
From this subsection, we consider the realistic model.  First, we take up the nanosheet with a thickness of $t$ 
as shown in Fig. 3.
\begin{figure}
\includegraphics{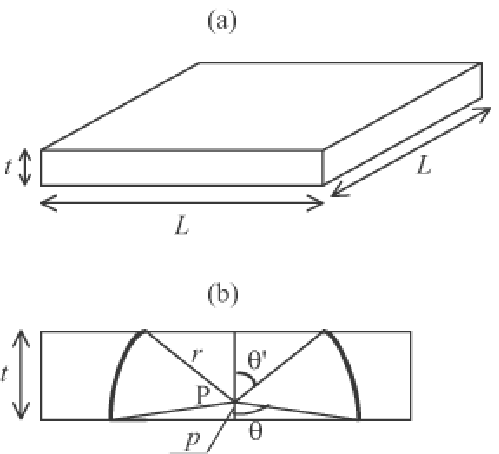} 
\caption{(a) Three dimensional scheme of the nanosheet with a thickness $t$. 
(b) The sectioned diagram of the nanosheet. }
\end{figure}
The area is $L \times L$ and it is assumed that $r \ll L$.  The atomic density in the sheet is $\rho_{0, 3D}'$.  
Here, we also consider the PRDF at $p$, $R(r, p)$, where $p$ is a distance 
between the center of the sphere with a radius $r$ and the bottom wall of the sheet, as shown in Fig. 3(b).
$R(r, p)$ is obtained as product of $\rho_{0, 3D}'$ and the volume of the overlapping part of the sheet and 
the spherical shell with a radius $r$ and a thickness $dr$, shown by the bold line in the figure.  
$R_{sheet}(r)$ is given by $\int_0^t R(r, p)dp/t$.  
In the case of $r \le t/2$, $R(r, p)$ is given as the number of atoms in the complete spherical 
shell for $r \le p \le t-r$.  
$R(r, p)$ is proportional to the surface area of the object obtained by rotating the fan 
with a interior angle $\theta$ and $\theta '$ 
for $p \le r$ and $t-r \le p \le t$, respectively, where $\theta=\cos^{-1} \frac{p}{r}$ 
and $\theta '=\cos^{-1} \frac{t-p}{r}$.  Then, for $r \le t/2$,
\begin{align}
R_{sheet}(r)&=\left[ \int_0^{r} 2\pi r^2 \left( 1+\frac{p}{r} \right) dp+\int_{r}^{t-r} 4\pi r^2 dp \right. \nonumber \\
 & \left. +\int_{t-r}^t 2\pi r^2 \left( 1+\frac{t-p}{r} \right) dp \right] \rho_{0, 3D}'/t  \nonumber.
\end{align}
In the case of $t/2 \le r \le t$, $R(r, p)=2\pi r^2 (1+p/r) \rho_{0, 3D}'$ for $p \le t-r$,  
and $R(r, p)=2\pi r^2 (1+\frac{t-p}{r}) \rho_{0, 3D}'$ for $r \le p \le t$.  For $t-r \le p \le r$, $R(r, p)$ is given as a 
product of $\rho_{0, 3D}'$ and the surface area of the object obtained by rotating the bold line shown in Fig. 3(b).  
For $t/2 \le r \le t$,
\begin{align}
R_{sheet}(r)&=\left[ \int_0^{t-r} 2\pi r^2 \left( 1+\frac{p}{r} \right) dp+\int_{t-r}^{r} 2\pi r t dp \right. \nonumber \\
 & \left. +\int_{r}^t 2\pi r^2 \left( 1+\frac{t-p}{r} \right) dp \right] \rho_{0, 3D}'/t  \nonumber.
\end{align}
Since in the case of $r \ge t$, $R(r, p)=2\pi r t$ for all $p$, 
\begin{equation}
R_{sheet}(r)=\int_{0}^{t} 2\pi r t dp \rho_{0, 3D}'/t  \nonumber.
\end{equation} 
By calculating the above integrations, we get the RDF of the nanosheet as follows.
\begin{align}
R_{sheet}(r) &=(4\pi r^2 t-2\pi r^3)\rho_{0, 3D}'/t \mspace{36mu} (r \le t) \nonumber \\
R_{sheet}(r) &=2\pi t r \rho_{0, 3D}' \mspace{36mu}  (r \ge t).
\end{align} 
\begin{figure}
\includegraphics{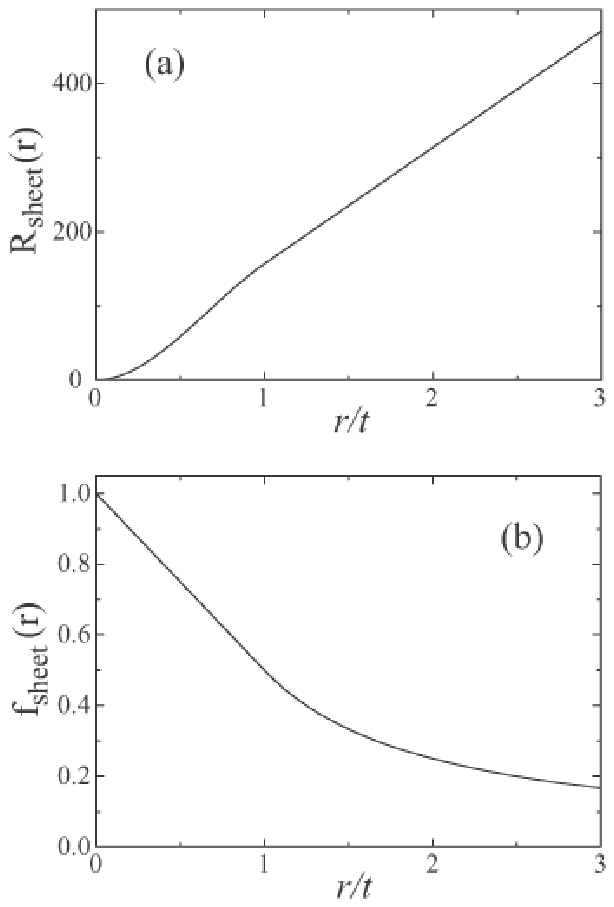} 
\caption{(a) The radial distribution function and (b) the correction factor 
of the nanosheet with thicknesses $t$.}
\end{figure}
Figure 4(a) shows the RDF calculated on the nanosheet.
In the calculations, the atomic density in the nanosheet $\rho_{0, 3D}'$ is unity.  For $r \ll t$, the RDFs 
are proportional to $r^2$.  This $r$-dependence corresponds with the case of three dimensional 
particle mentioned in \S 2.  For $r > t$, $R_{sheet}(r) \propto r$, corresponding with $R_{2D}(r)$.  
The correction factor $f_{sheet}(r)$ which is defined by eq. (\ref{f(r)}) 
is shown in Fig. 4(b).  At $r=0$, $f(r)$ is unity, and it linearly decreases with $r$ in the region 
$0 < r < t$.

\subsection{Nanobelt}
Next, we consider the case of the nanobelt with a width $a$ and a thickness $t$, as shown in Fig. 5.
\begin{figure}
\includegraphics{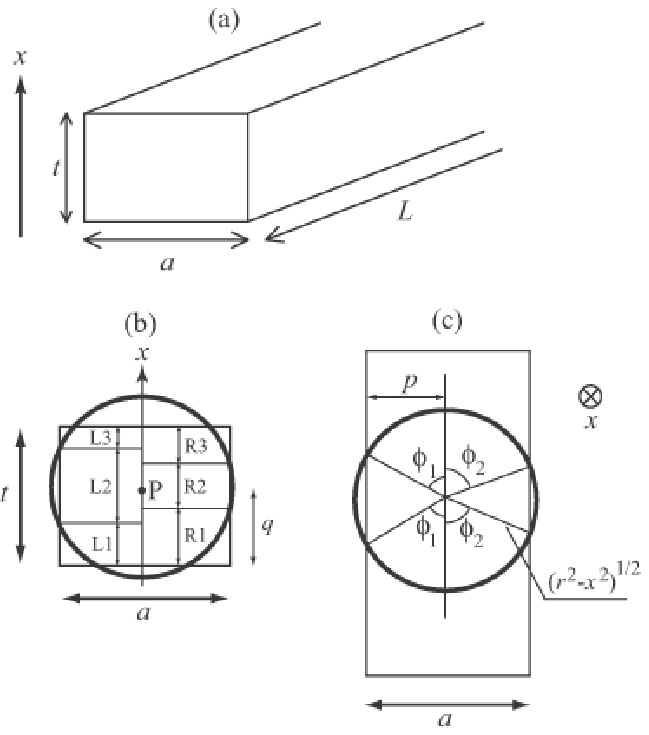} 
\caption{(a) Three dimensional scheme of nanobelt with width $a$ and thickness $t$. 
(b) The sectioned diagram of the nanobelt.  (c) The diagram viewed along $x$-direction. }
\end{figure}
The length of the belt $L$ is much larger than $r$.  Figure 5(b) shows a sectioned diagram of the belt.  The center of 
the circle (P) in the figure is distant by $p$ from the left wall and by $q$ from the bottom wall.  
The PRDF at $(p, q)$, 
$R(r, p, q)$, is proportional to the surface area of the part of the sphere which cross the belt.  The surface area 
of the object obtained by rotating by $\phi(x)$  around $x$ axis with the radius $y(x)$ is obtained by 
$S=\int \int_{\phi(x)} y(x) d\phi ds$, where $ds=\sqrt{dx^2+dy^2}$.  Since in the case of sphere, $y=\sqrt{r^2-x^2}$, 
$S=\int r \phi(x) dx$.  Here, we calculate $R(r, p, q)$ by separating six parts as shown in Fig. 5(b).  
The surface area of L2 is obtained by 
the rotation from zero to $\phi_1(x)$ and from $\pi-\phi_1(x)$ to $\pi$, and the area of R2 is obtained by the 
rotation from zero to $\phi_2(x)$ and from $\pi-\phi_2(x)$ to $\pi$, as shown in Fig. 5(c).  The surface areas of 
the other parts are obtained by the rotation of $\pi$.  Then
$R(r, p, q)$ is given by
\begin{align}
R(r, p, q)=& \rho_{0, 3D}' \left[ \int_{L1} \pi r dx+\int_{L2} 2\phi_{1}(x) r dx +\int_{L3} \pi r dx \right.  \nonumber \\
             &\left. +\int_{R1} \pi r dx+\int_{R2} 2\phi_{2}(x) r dx +\int_{R3} \pi r dx \right] . \label{Rrpq}
\end{align} 
The direction of $x$-axis is shown in Fig. 5(b), and the origin of $x$-axis corresponds 
with the center of the sphere.  $\phi_{1}(x)$ and $\phi_{2}(x)$ are given by the relations, 
\begin{align}
 \phi_{1}(x) &=\sin^{-1} \frac{p} {\sqrt{r^2-x^2}}, \nonumber \\
 \phi_{2}(x) &=\sin^{-1} \frac {a-p} {\sqrt{r^2-x^2}}, \nonumber
\end{align} 
as shown in Fig. 5(c).  The integral ranges in eq. (\ref{Rrpq}) depend on the relations 
between $r$, $t$, $a$, $p$ and $q$.  
The integral ranges of the left side (L1, L2 and L3) 
for various conditions of the above parameters are shown in Tabel 1.
\begin{table}
\begin{center}
    \caption{Integral ranges of the left side (L1, L2 and L3) 
for various conditions of the relations between $r$, $t$, $a$, $p$ and $q$.}
    \label{table:1}
    \begin{tabular}{llll}
    conditions & L1 & L2 & L3 \\ \hline
     (i) $r\le p$ &  & & \\
     $r\le q\le t-r$ & $\int_{-r}^rdx$ &   &   \\
       $q\le t-r$,  $q\ge r$ & $\int_{-r}^{t-q}dx$ &   &   \\
         $q\le t-r$,  $q\le r$ & $\int_{-q}^rdx$ &   &   \\
       $t-r\le q\le r$ & $\int_{-q}^{t-q}dx$ &   &   \\ \hline
     (ii) $r>p$ &  & & \\
   $t-\sqrt{r^2-p^2}<q<\sqrt{r^2-p^2}$ &   & $\int_{-q}^{t-q}dx$ &   \\
        $\sqrt{r^2-p^2} < q < r$ and, &&& \\ 
        $q>t-\sqrt{r^2-p^2}$ & $\int_{-q}^{-\sqrt{r^2-p^2}}dx$ & $\int_{-\sqrt{r^2-p^2}}^{t-q}dx$ &   \\
        $q>r$, $q>t-\sqrt{r^2-p^2}$ & $\int_{-r}^ {-\sqrt{r^2-p^2}}dx$ & $\int_{-\sqrt{r^2-p^2}}^{t-q}$ &   \\
        $q<\sqrt{r^2-p^2}$ and, &&& \\
        $q<t-r$ &    & $\int_{-q}^{\sqrt{r^2-p^2}}dx$ & $\int_{\sqrt{r^2-p^2}}^rdx$ \\
        $q<\sqrt{r^2-p^2}$ and, &&&\\
        $t-q \le q \le t-\sqrt{r^2-p^2}$ &   & $\int_{-q}^{\sqrt{r^2-p^2}}dx$ & $\int_{\sqrt{r^2-p^2}}^{t-q}dx$ \\
        $r<q<t-r$ & $\int_{-r}^{-\sqrt{r^2-p^2}}dx$ & $\int_{-\sqrt{r^2-p^2}}^{\sqrt{r^2-p^2}}dx$ & $\int_{\sqrt{r^2-p^2}}^rdx$ \\
        $q>r$ and, &&&\\ 
        $t-r<q<t-\sqrt{r^2-p^2}$ & $\int_{-r}^{-\sqrt{r^2-p^2}}dx$ & $\int_{-\sqrt{r^2-p^2}}^{\sqrt{r^2-p^2}}dx$ & $\int_{\sqrt{r^2-p^2}}^{t-q}$ \\
        $\sqrt{r^2-p^2}<q<r$, and &&& \\ 
        $q<t-r$ & $\int_{-q}^{-\sqrt{r^2-p^2}}dx$ & $\int_{-\sqrt{r^2-p^2}}^{\sqrt{r^2-p^2}}dx$ & $\int_{\sqrt{r^2-p^2}}^rdx$ \\
        $\sqrt{r^2-p^2}<q<r$ and, &&&\\
        $t-r<q<t-\sqrt{r^2-p^2}$ & $\int_{-q}^{-\sqrt{r^2-p^2}}dx$ & $\int_{-\sqrt{r^2-p^2}}^{\sqrt{r^2-p^2}}dx$ & $\int_{\sqrt{r^2-p^2}}^{t-q}dx$  \\ \hline
 \end{tabular}
 \end{center}
 \end{table}
 For R1, R2 and R3, the integral ranges are obtained by substituting $a-p$ for $p$ in the table.  The case shown 
 in Fig. 5(b) corresponds with the condition of $r>p$, $\sqrt{r^2-p^2}<q<r$ and $t-r<q<t-\sqrt{r^2-p^2}$.
Because the integrations for the range L2 and R2 can not be calculated analytically, they are obtained by 
the numerical calculations.
$R_{belt}(r)$ is given by,
\begin{equation}
R_{belt}(r)=\int_{0}^{a} \int_{0}^{t} R(r, p, q)dqdp/(ta)  \nonumber.
\end{equation} 
The RDFs calculated for the nanobelts with various thickness $t$ are shown in Fig. 6(a).
\begin{figure}
\includegraphics{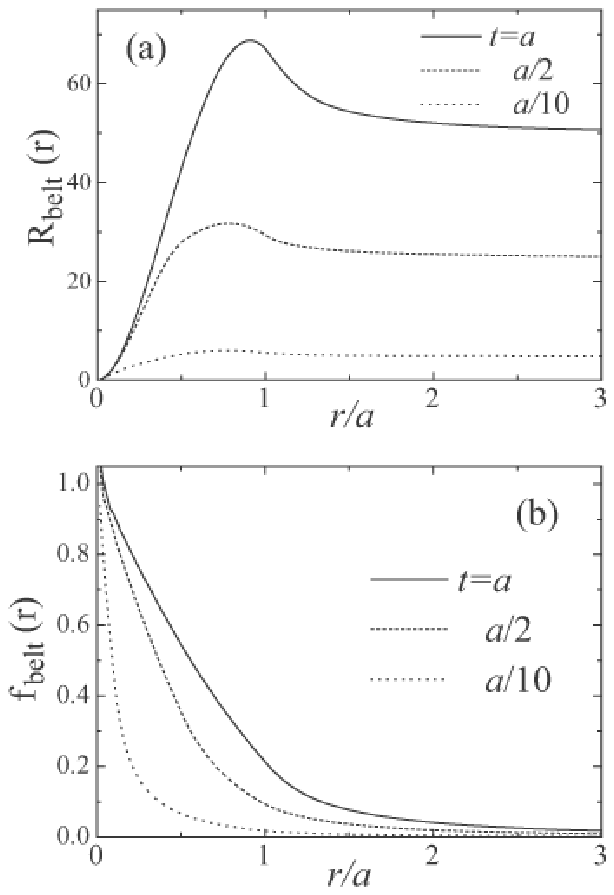} 
\caption{(a) The radial distribution functions and (b) the correction factors 
of the nanobelts with various $t$. }
\end{figure}
For $r \gg a, t$, $R_{belt}(r)$ become flat and they are proportional 
to $at\rho_{0, 3D}'$, corresponding with $R_{1D}(r)$.  The correction factors obtained by 
eq. (\ref{f(r)}) are shown in Fig. 6(b).  
At $r \sim 0$, $f_{belt}(r)$ for $t=a$ and $t=a/2$ are slighty larger than unity, due to the insufficiency of the 
accuracy of the numerical calculation.\par

\subsection{Nanorod and Nanotube}

In this subsection, first, we consider the nanorod as shown in Fig. 7(a).
\begin{figure}
\includegraphics{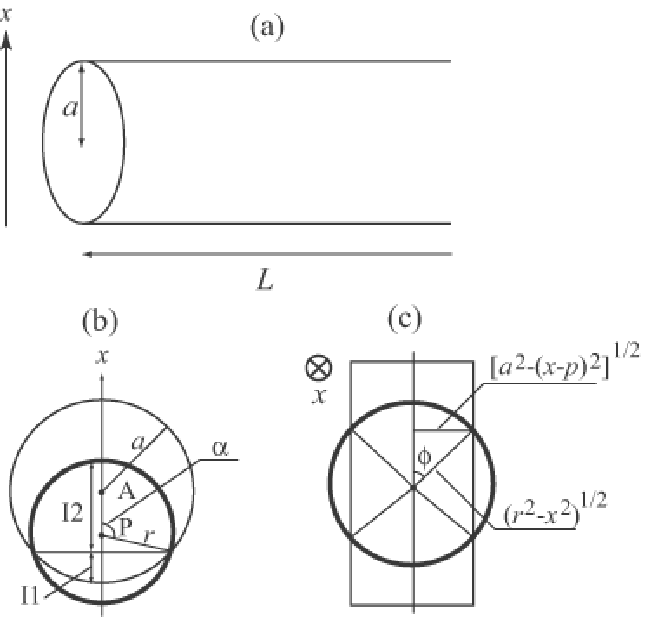} 
\caption{(a) Nanorod with radius $a$. (b) The sectioned diagram of the nanorod whose center is located on A. 
(c) The diagram viewed along $x$-direction.}
\end{figure}
We also assume that $r \ll L$.  Figure 7(b) shows a sectioned diagram.  We consider the surface area of 
the overlapping part of 
the sphere with a radius $r$ and the rod with a radius $a$.  
The center of the sphere is distant from the center of the rod 
by distance PA=$p$.  The PRDF at $p$, $R(r, p)$ is proportional to the surface area.
It is similar to the cases of the nanosheet and the nanobelt.  $R(r, p)$ is obtained as
\begin{equation}
R(r, p)= \left[ \int_{I1} 4\phi (x) r dx+\int_{I2} 2\pi r dx \right] \rho_{0, 3D}', \nonumber
\end{equation} 
where $\phi (x)$ is given by
\begin{equation}
 \phi (x) =\sin^{-1}\sqrt{\frac{a^2-(x-p)^2}{r^2-x^2}}, \nonumber \\
\end{equation}
as shown in Fig. 7(c).  The integral ranges I1 and I2 depend on the relation between the parameters, $a$, $r$ and $p$.
First, the case of $r<a$ is considered.  In this case, when $p-a<-r$, the sphere is perfectly enveloped in the rod.  
If $p-a>-r$, the integral ranges on $x$ are obtained as, I1:$p-a \le x \le r\cos\alpha$ and 
I2:$r\cos\alpha \le x \le r$, where $\cos\alpha=\frac{p^2+r^2-a^2}{2pr}$.  Then $R(r, p)$ is given for $r<a$,
\begin{align}
R(r, p)&=4\pi r^2 \rho_{0, 3D}' \mspace{36mu} (p \le a-r)  \nonumber \\
R(r, p)&=\left[ \int_{p-a}^{r\cos\alpha} 4\phi (x)r dx + \int_{r\cos\alpha}^{r} 2\pi r dx \right] \rho_{0, 3D}' \nonumber\\
       &=\left[ 4r\int_{p-a}^{r\cos\alpha}\sin^{-1}\sqrt{\frac{a^2-(x-p)^2}{r^2-x^2}} dx \right. \nonumber \\
       &\left. +2\pi r^2 \left( 1-\frac{p^2+r^2-a^2}{2pr} \right) \right] \rho_{0, 3D}'  \mspace{36mu} (p\ge a-r). \label{Rrprod1}
\end{align}
Since the RDF is given by $\int_0^a R(r, p) 2\pi pdp/\pi a^2$, $R_{rod}(r)$ is obtained from the above 
equations as
\begin{align}
R_{rod}(r)&=\rho_{0, 3D}' \int_{0}^{a-r} 4\pi r^2 \times 2\pi p dp /\pi a^2  \nonumber\\
          &+\rho_{0, 3D}' \int_{a-r}^a \left[ 2\pi r^2 \left( 1-\frac{p^2+r^2-a^2}{2pr} \right) 2\pi p \right. \nonumber\\
          &\left. +4r\int_{p-a}^{r\cos\alpha}\sin^{-1}\sqrt{\frac{a^2-(x-p)^2}{r^2-x^2}} dx \right] \nonumber \\
          & \times 2\pi pdp /\pi a^2 \mspace{36mu} (r<a). \label{Rrod}
\end{align}
The integration on $x$ must be obtained by the numerical calculation.\par
In the case that $r>a$, $R(r, p)$ can be given by the second formula in eq. (\ref{Rrprod1}) when $p+a>r$.  
When $p+a<r$, the circle with a radius $a$ 
which corresponds with a section of the rod, is completely enveloped by the circle with a radius $r$.  Then $R(r, p)$ 
is given by
\begin{align}
R(r, p)&=\rho_{0, 3D}' \int_{p-a}^{p+a} 4\phi (x) r dx \nonumber\\
       &=4r \rho_{0, 3D}' \int_{p-a}^{p+a} \sin^{-1}\sqrt{\frac{a^2-(x-p)^2}{r^2-x^2}} dx.
\end{align}
As the results, for $r>a$, $R_{rod}$ is obtained as
\begin{align}
R_{rod}(r)&=\rho_{0, 3D}' \int_{0}^{r-a} \int_{p-a}^{p+a} 4r \sin^{-1}\sqrt{\frac{a^2-(x-p)^2}{r^2-x^2}} dx \nonumber \\
          & \times  2\pi pdp /\pi a^2 \nonumber \\
          & + \rho_{0, 3D}' \int_{r-a}^a \left[  2\pi r^2 \left( 1-\frac{p^2+r^2-a^2}{2pr} \right) \right. \nonumber \\
          & +  \left. 4r\int_{p-a}^{r\cos\alpha}\sin^{-1}\sqrt{\frac{a^2-(x-p)^2}{r^2-x^2}} dx \right]  \nonumber \\
          & \times  2\pi pdp /\pi a^2 \mspace{36mu} (r\ge a).
\end{align}
The integrations on $x$ included in the first and the third terms are calculated numerically. \par
Next, we consider the nanotube shown in Fig. 8.  
\begin{figure}
\includegraphics{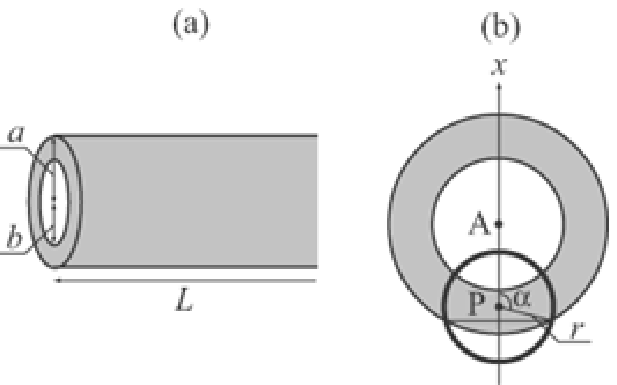} 
\caption{(a) Nanotube with an external diameter 2$a$ and an internal diameter 2$b$.  
(b) Sectioned diagram of the nanotube. }
\end{figure}
An external diameter and an internal diameter are 2$a$ and 2$b$, respectively.  In the calculation, 
we consider the PRDFs at $p$ for the nanorods with radii $a$ and $b$, $R_a(r, p)$ and $R_b(r, p)$, where 
$p$ is a distance between the center of the tube and the sphere with a radius $r$.  The RDF 
of the nanotube is given by using $R_a(r, p)$ and $R_b(r, p)$, as follows.
\begin{equation}
R_{tube}(r)=\int_b^a \left[ R_a(r, p) - R_b(r, p) \right]  2\pi p dp/\pi (a^2-b^2).
\end{equation}
$R_a(r, p)$ can be calculated by same way as the case of the rod.  $R_b(r, p)$ can be given for the conditions of the 
parameters,
\begin{align}
R_b(r, p)&=4r \rho_{0, 3D}' \int_{p-b}^{p+b}\sin^{-1}\sqrt{\frac{b^2-(x-p)^2}{r^2-x^2}}dx \nonumber \\
 &(p \le r-b), \nonumber \\
R_b(r, p)&=\rho_{0, 3D}' \left[ 2 \pi r^2 \left( 1-\frac{p^2+r^2-b^2}{2pr} \right) \right. \nonumber \\
         & \left.  +4r\int_{r\cos\alpha}^{p-b}\sin^{-1}\sqrt{\frac{b^2-(x-p)^2}{r^2-x^2}}dx \right] 
\mspace{18mu} (p \ge r-b), \nonumber \\
R_b(r, p)&=0 \mspace{36mu} (p \ge r+b).         
\end{align}
By using the above equations, we calculate the RDFs of the nanorod and the nanotubes, as shown in Fig. 9(a).
\begin{figure}
\includegraphics{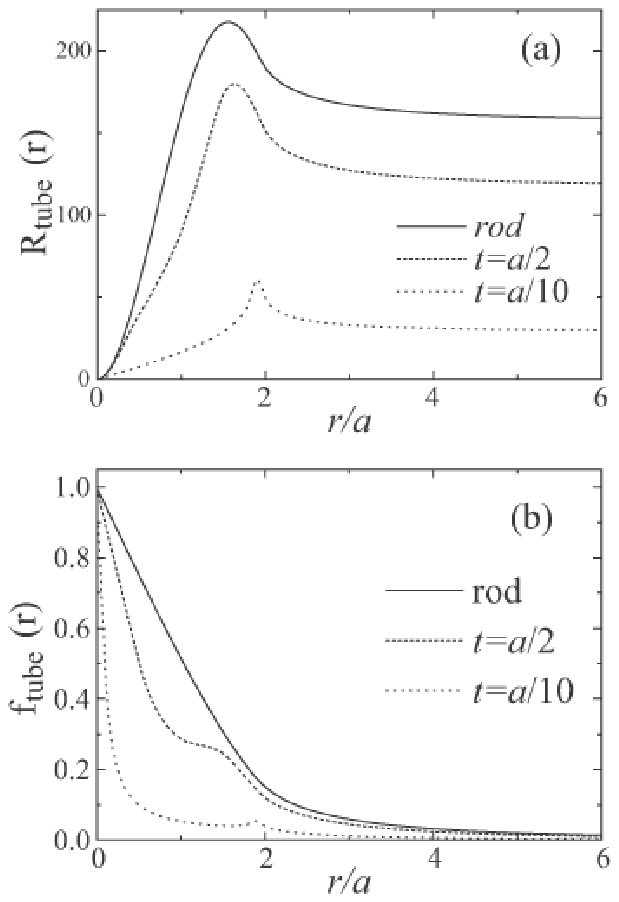}
\caption{(a) The radial distribution functions and (b) the correction factors 
of the nanorod and the nanotubes with various $t$. }
\end{figure}
In these figures, a thickness $t=a-b$.  The atomic densities in the rod and the tube, $\rho_{3D}'$ are unity.
At $r \gg a, t$, the RDFs have constant values which correspond with 
the a product of the sectioned area and the atomic density.  It is same as the case of the nanobelt.  
$R_{tube}(r)$ with a thin thickness have sharp peaks at $r \sim 2a$.  
The correction factors $f_{tube}(r)$ are shown in Fig. 9(b).  
The correction factor of the tube decreases with $r$, and has a shoulder at $r\sim 2a-t$. \par

\subsection{Sphere}
In this subsection, the RDFs for the sphere and the spherical shell are calculated.  
Fisrt, the filled sphere with a radius $a$ 
as shown in Fig. 10(a) is considered.  
\begin{figure}
\includegraphics{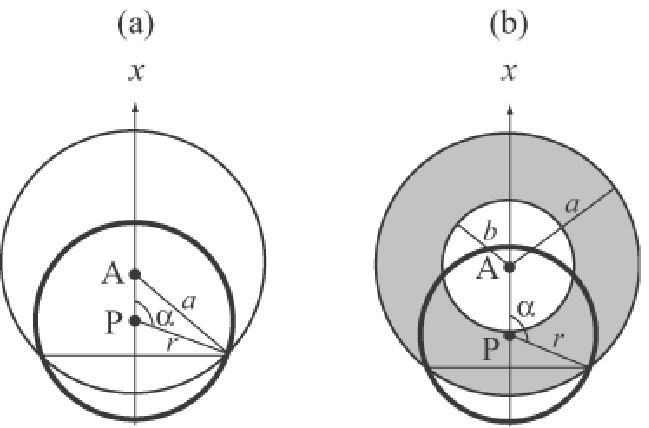} 
\caption{Sectioned diagram of (a) filled sphere and (b) spherical shell 
with external and internal radii $a$ and $b$, respectively. }
\end{figure}
In this case, we also consider $R(r, p)$ where $p$ is a length PA shown in Fig. 10.  
In the case of $r<a$, it can be obtained as
\begin{align}
R(r, p)&=4\pi r^2 \rho_{0, 3D}' \mspace{18mu} (p \le a-r)  \nonumber \\    
R(r, p)& =\rho_{0, 3D}'\int_{r\cos\alpha}^r 2\pi r dx \nonumber \\
       & =2\pi r^2 \rho_{0, 3D}'\left( 1-\frac{p^2+r^2-a^2}{2pr} \right)  \mspace{18mu} (p \ge a-r) . \label{sphere1}
\end{align}       
In the case of $a<r<2a$,
\begin{align}
R(r, p)&=0  \mspace{18mu} (p \le r-a)  \nonumber \\    
R(r, p)& =\rho_{0, 3D}'\int_{r\cos\alpha}^r 2\pi r dx \nonumber \\
       & =2\rho_{0, 3D}' \pi r^2 \left( 1-\frac{p^2+r^2-a^2}{2pr} \right)  \mspace{18mu} (p \ge r-a) . \label{sphere2}
\end{align}     
For $r>2a$, $R(r, p)=0$.  
The RDF can be given by
\begin{equation}
R_{Fsphere}(r)=\frac{\int_0^a R(r, p) 4\pi p^2 dp}{\frac{4}{3}\pi a^3}
\end{equation}
The above integrations of eqs. (\ref{sphere1}) and (\ref{sphere2}) give same results.  Then
\begin{align}
R_{Fsphere}(r)&=\pi r^2 \rho_{0, 3D}' \left[ \frac{1}{4} \left( \frac{r}{a} \right) ^3 -3 \frac{r}{a} +4 \right] \mspace{18mu} (r \le 2a), \nonumber \\
R_{Fsphere}(r)&=0 \mspace{18mu} (r > 2a).
\end{align}
It is consistent with the result obtained by Mason.(1968) \par
In the case of the spherical shell with a thickness $t=a-b$ shown in Fig. 10(b), we consider the PRDF at $p$ 
for the filled spheres 
with radii $a$ and $b$, $R_a(r, p)$ and $R_b(r, p)$, respectively, and the RDF can be given by  
\begin{equation}
R_{Esphere}(r)=\frac{\int_b^a \left[ R_a(r, p)-R_b(r, p) \right]  4\pi p^2dp}{\frac{4}{3} \pi \left( a^3-b^3 \right)}. \label{sphere3}
\end{equation}
$R_a(r, p)$ corresponds with eqs. (\ref{sphere1}) and (\ref{sphere2}).  $R_b(r, p)$ is obtained as
\begin{align}
R_b(r, p)&=0 \mspace{18mu} (p < r-b) \nonumber \\
R_b(r, p)&=2 \pi r^2 \rho_{0, 3D}' \left( 1- \frac{p^2+r^2-b^2}{2pr} \right) \nonumber \\ 
&\mspace{18mu} (r-b \le p \le r+b)\nonumber \\
R_b(r, p)&=0 \mspace{18mu} (p > r+b) \label{sphere4}
\end{align}
By using eqs. (\ref{sphere1}), (\ref{sphere2}), (\ref{sphere3}) and (\ref{sphere4}), the RDFs 
for the spherical shells with various 
$t$ are calculated by using $\rho_{0, 3D}=1$, as shown in Fig. 11(a).  
The correction factors are shown in Fig. 11(c).
\begin{figure}
\includegraphics{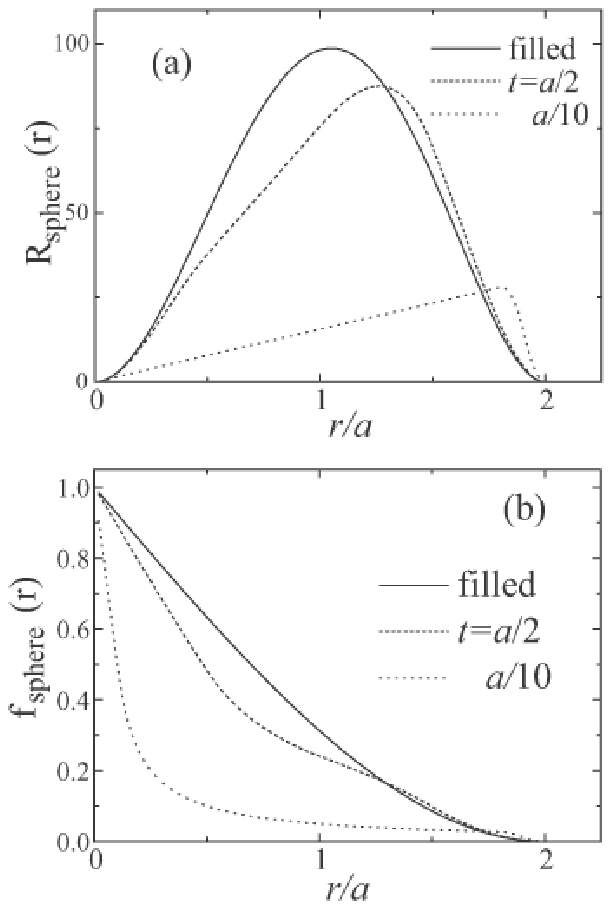} 
\caption{(a) The radial distribution functions and (b) the correction factors 
of the nanospheres with various $t$. }
\end{figure}

\section{Discussions}

In this section, we present the formulation of the PDF analysis 
which takes account of the finite size effect of the nanoparticle.
In the previous section, we neglect the atomic pair 
distribution between the particles, however it must be considered in the real analysis.  
Here, we consider the partial raidal distribution function including the atomic pair distributions inside of a particle 
and between the particles.  
It is assumed that the measured nanoparticles are in a vacuum (there is not any scatterer or atom outside of particles).  
Because in the real case, each particle is usually distributed randomly 
and there may be no atomic correlation between particles, 
the atomic distribution outside of the particle can be regarded as being continuous with a density 
of $\rho_0$ which corresponds with the density averaged in the whole sample including all particles.  
The atomic arrangement inside of particles is periodic with a density of $\rho_0'$.
Here we rewrite the radial distribution and the partial radial distribution functions obtained 
for the single particle made of atoms as $R_{nano}(r)$ and $R_{nano}(r, p)$ (or $R_{nano}(r, p, q)$), 
respectively.
Then the partial radial distribution function $R(r, p)$ (or $R(r, p, q)$) 
are given as follows.   
\begin{equation}
R(r, p) =R_{nano}(r, p)+\left[4\pi r^2-S_{nano}(r, p) \right] \rho_0 \nonumber  
\end{equation}
where $S_{nano}(r, p)$ shows a surface area  
that the sphere with a radius $r$ centered at P has inside of the particle.  
In the above equation, the first term shows the pair distribution inside of the particle and the second term shows 
the pair distribution of outside of the particle (the pair distribution between the particles).  
If the nanoparticle consists of continua, $R_{nano}(r, p)/\rho_0'$ becomes equal to $S_{nano}(r, p)$.
Then the radial distribution function is given by
\begin{equation}
R(r) =R_{nano}(r)+\left[ 4\pi r^2-S_{nano}(r) \right] \rho_0   \nonumber , 
\end{equation}
where $S_{nano}(r)$ is the surface area inside of the particle averaged by $p$.  
In the previous section, we know that $4\pi r^2f(r)$ is substituted for $S_{nano}(r)$.   
Then the RDF can be represented as
\begin{equation}
R(r) =f(r) R_{\infty}(r) +4\pi r^2 \rho_0 \left[ 1-f(r) \right]    \label{Rnano}, 
\end{equation}
where $R_{\infty}(r)$ is RDF of the bulk sample with infinite size.
Here we use the relation
\begin{equation}
f(r) =R_{nano}(r)/ R_{\infty}(r),  \nonumber
\end{equation}
for the first term.  
For the bulk sample, the PDF and the reduced PDF are given by 
\begin{align}
g_{\infty}(r)&=R_{\infty}(r)/4\pi r^2 \rho_0'  \nonumber \\
G_{\infty}(r)&=4\pi r \rho_0' [g_{\infty}(r)-1] \label{bulk} ,
\end{align}
where $g_{\infty}(r)$ and $G_{\infty}(r)$ are the PDF and the reduced PDF of the bulk sample.
From eqs. (\ref{g(r)}), (\ref{G(r)}), (\ref{Rnano}) and (\ref{bulk}), the PDF and the reduced PDF 
are represented as
\begin{align}
g(r) &= \frac{\rho _{0}'} {\rho _{0}} f(r) g_{\infty}(r) +1-f(r), \label{gnano} \\ 
G(r)&=  f(r) G_{\infty}(r) +4\pi r ( \rho _{0}' -\rho_{0} ) f(r) \label{Gnano} .
\end{align} 
The RDF, the PDF and the reduced PDF of the nanoparticle are obtained from 
those of the bulk sample modified by $f(r)$ and the ratio of $\rho_{0}'/\rho_{0}$.  
The above $g(r)$ and $G(r)$ satisfy the normalization relations of $g(r) \rightarrow 1$ 
and $G(r) \rightarrow 0$ in the limit of $r \rightarrow \infty$ because $f(r)\rightarrow 0$ for $r \rightarrow \infty$.  
Then the eqs. (\ref{Rnano}), (\ref{gnano}) and (\ref{Gnano}) are valid, and  
they can be applied to the PDF analysis on nanoparticles. \par
In eq. (\ref{Gnano}), the second term is independent of the arrangement of the atoms included in the nanoparticle, 
and it depends only on the shape and the size of the nanoparticle (and the densities), 
while the first term depends on both the atomic arrangement
in the nanoparticle, and the shape and the size of the nanoparticle.
As a result, in the general case, we can expect that the peak structure due to 
the atomic arrangement given by the first term 
is on the "back ground" due to the shape of the nanoparticle given by the second term which can be easily calculated by 
using $f(r)$.  From eq. (\ref{Gnano}), the "back ground" is larger for smaller $\rho_{0}$. 
Then the detailed experiment on the diluted nanoparticle can confirm the above expectations. \par
Here, we discuss the difference between the PDF analysis and the small angle scattering.  
Generally, the shape and the size of the nanoparticles are determined by the small angle scattering, and the scattering 
functions for the particles with various shapes have been calculated.  However, since in the small 
angle scattering measurement, the wave vector is much smaller than the inverse of the atomic scale 
and the particle can be regarded as a 
continuum, the scattering profile depends on the shape and the size of 
the particle and is independent of the atomic arrangement.  It indicates that the small angle scattering 
can not distinguish the local disordered lattice from the ordered lattice, 
although, generally, they coexist in the nanoparticle.  On the other hand, 
the PDF analysis can distinguish the local disordered lattice from the ordered lattice, 
because the functions of the PDF analysis depend on the atomic arrangement and the coherence.  Then the PDF analysis 
can discuss the local lattice disorder in the nanoparticle by considering the finite size effect of the nanoparticle.   
Gilbert \textit{et al.} (2004) have shown that the PDF profile of the zinc sulfide nanoparticle decreases with $r$ 
more rapidly than the profile expected from the shape and the size of the particle.  They pointed out that 
such reduction of the PDF profile is due to the local structural disorder driven by the strain in the particle 
and estimate the distance in which the structural coherence remains, from the reduction of the PDF profile.  
They speculate that such strain is caused by the irregular surface.\par
Here, we can present a method to estimate the sizes of the domains with the ordered and the disordered lattice 
in nanoparticles, based on the present calculations.  
In the PDF analysis, the atomic density in the disordered part 
becomes nearly a continuum.  For the continuum, $G_{\infty}(r)$ is zero, which can be obtained from the relation 
$R_{\infty}(r)=4\pi r^2 \rho_0'$, and eq. (\ref{bulk}).  
Then the reduced PDF of continuum with nano-size is given by the smooth curve 
which corresponds with the second term of eq. (\ref{Gnano}).  
The reduced PDF of the part with the structural coherence 
shows sharp peaks corresponding to the atomic arrangement.   
Let's consider a spherical nanoparticle with a disordered surface, as a simple example.
In this case, we consider two part : the sphere with a periodic atomic arrangement which 
has a diameter smaller than that of the particle, and 
the spherical shell of a continuum as shown in Fig. 10(b).  
The reduced PDF of the former are represented by modifying the functions of the bulk sample by using eq. 
(\ref{Gnano}).  $G(r)$ of the latter is given by the second term of eq. (\ref{Gnano}), obtained by using $f(r)$ of 
the spherical shell with a corresponding size.  We may calculate a sum of the two kinds of the reduced PDFs and fit it to 
the observed data by adjusting the thickness of the disordered surface, and as a result, 
we can estimate thickness of the disordered surface.  
By using our calculations, the information on the local inhomogenity in nanoparticles may be detected quantitatively.\par
  
\section{Conclusion}

The functions of the PDF analysis calculated for the various nanoparticles strongly depend on their shapes and sizes.  
The exact equations of the PDF analysis on the nanoparticle are presented, 
by considering the correction factor $f(r)$ and the ratio between the density 
in the particle and the total density averaged in the sample.  The analysis which takes account of 
the finite size effect also enables us to estimate the sizes of the local 
disordered lattice and the ordered lattice included in the nanoparticle. \par
The finite size effect on the PDF is remarkable in the large $r$-region 
where the accurate experimental data is hard to be obtained.  
The detailed analysis is enabled by a diffractometer 
with both of high resolution and high intensity, 
which can be installed only in intense pulse neutron or intense synchrotron radiation source facilities. 




\ack{Acknowledgements}

This work was supported by a Grant-in-Aid for Scientific Research from the Ministry of Education, 
Culture, Sports, Science and Technology of Japan.




\end{document}